\documentstyle[ApJ,times,psfig]{article}

\def\spose#1{\hbox to 0pt{#1\hss}}
\def\lta{\mathrel{\spose{\lower 3pt\hbox{$\mathchar"218$}}
     \raise 2.0pt\hbox{$\mathchar"13C$}}}
\def\gta{\mathrel{\spose{\lower 3pt\hbox{$\mathchar"218$}}
     \raise 2.0pt\hbox{$\mathchar"13E$}}}
\def\Meszaros{M\'esz\'aros~}

\begin{document}

\title{Steepening of Afterglow Decay for Jets Interacting with Stratified Media}
\author{Pawan Kumar}
\affil{IAS, Princeton, NJ 08540}
\author{Alin Panaitescu}
\affil{Dept. of Astrophysical Sciences, Princeton University, NJ 08544}
\authoremail{pk@ias.edu, adp@astro.princeton.edu}

\begin{abstract}

We calculate light-curves for Gamma-Ray Burst afterglows when
material ejected in the explosion is confined to a jet
which propagates in a medium with a power-law density profile.
The observed light-curve decay steepens by a factor of $\Gamma^2$ when
an observer sees the edge of the jet.
In a uniform density medium the increase in the power-law index ($\beta$) 
of the light-curve as a result of this {\it edge effect} is $\sim0.7$ and is 
completed over one decade in observer time. For a pre-ejected stellar wind 
($\rho \propto r^{-2}$) $\beta$ increases by $\sim0.4$ over two decades in time
due to the edge effect and the steepening of the light-curve due to the 
jet sideways expansion takes about four decades in time.
Therefore, a break in the light-curve for a jet in a wind model is unlikely 
to be detected even for very narrow jets of opening angle of a few degrees 
or less, in which case the lateral expansion occurs at early times when the 
afterglow is bright.

The light-curve for the afterglow of GRB 990510, for which an increase 
in $\beta$ of approximately 1.35 was observed on a time scale of 3 days, 
cannot be explained only by the sideways expansion and the edge effects 
in a jet in a uniform ISM -- the increase in $\beta$ is too large and too 
rapid. However, the passage of the cooling or synchrotron peak frequencies
through the observing band at about 0.1 -- 1 day together with jet edge 
effect explains the observed data. The jet opening angle is found to be
$\sim 5^o$ and the energy in the explosion to be less than about $10^{50}$ erg.

\end{abstract}

\keywords{gamma rays: bursts -- gamma-rays: theory}

\section{Introduction}

In a recent paper Chevalier and Li (1999) pointed out that
some of the GRB afterglow light-curves are best modeled when the
density of the circum-burst medium is taken to fall off as $r^{-2}$ 
(this is referred to as the {\it wind model}).
These afterglows show no evidence for a jet, i.e. their light-curves follow 
a power-law decline without any break. This is puzzling since collimated
outflows are expected in the collapsar model for GRBs (MacFadyen, Woosley \& 
Heger 2000). We offer a possible explanation for this puzzle by showing
that the light-curve resulting from the interaction of a
jet with a pre-ejected wind falls off as a power-law whose index
changes very slowly with time.

We carry out a detailed modeling of the multi-wavelength afterglow flux data 
for GRB 990510, which provides the best evidence for a jet propagation in 
a uniform density medium (Harrison et al. 1999, Stanek et al. 1999), to show 
that effects associated with a finite jet opening-angle are insufficient to 
explain the observed rapid steepening of the light-curve.

In \S\ref{dynamics} we calculate the propagation of a jet in a 
stratified medium and in \S\ref{synchrotron} we describe the calculation 
of the synchrotron emission and afterglow light-curve.

\section{Dynamics of Expanding Jets}
\label{dynamics}

The dynamical evolution of jets and its synchrotron emission have 
been previously investigated by a number of people, e.g. Rhoads (1999),
Panaitescu \& \Meszaros (1999), Sari, Piran \& Halpern (1999), Moderski, Sikora, \&
Bulik (2000), Huang et al. (2000). The evolution of the Lorentz factor ($\Gamma$)
can be calculated from the following set of equations
\begin{equation}
 {d M_1\over dr} = 2\pi A r^{2-s}(1-\cos\theta) \;,
\end{equation}
\begin{equation}
    {d \theta\over dr} = {1\over f r (\Gamma^2-1)^{1/2}} + {\Theta-\theta\over r} \;,
\label{theta}
\end{equation}
\begin{equation}
     M_0\Gamma + M_1(\Gamma^2-1) = M_0\Gamma_0 \;,
\end{equation}
where $\theta$ is the half-opening angle of the jet, $M_0$, and $\Gamma_0$ are 
the initial mass and Lorentz factor of the ejecta, $M_1$ is the swept-up mass, 
$\rho(r)=A r^{-s}$ is the density of the circum-stellar medium, and $f=c/c_s$ is
the ratio of the speed of light to that of the jet sideways expansion; $f$ is a
parameter of order unity whose effect can be absorbed in $\Gamma_0$, and 
which has little effect on the light-curve.
$\Theta$ is the angle between the velocity vector at the jet edge and
the jet axis (in the lab frame) and is determined by the modification of
particle trajectory due to the sideways expansion. The last equation above 
expresses the conservation of energy and it applies 
to an adiabatic shock when the heating of the original baryonic material
of rest mass $M_0$ by the reverse shock is ignored. 

The above equations can be combined and rewritten in the following 
non-dimensional form which is applicable for relativistic as well 
as non-relativistic jet dynamics 
\begin{equation}
{d y_1\over dx} = -{x^{2-s} (y_1^2-\Gamma_0^{-2})^2\, y_2^2\over 2y_1-y_1^2
   -\Gamma_0^{-2}} \;,  
\label{dynamic1}
\end{equation}
\begin{equation}
    {d y_2\over dx} = {1\over f \theta_0\Gamma_0 x(y_1^2-\Gamma_0^{-2})^{1/2}}
     + {\Theta-\theta\over x\theta_0},
\label{dynamic2}
\end{equation}
where $x = r/R_{da}$, $y_1 = \Gamma/\Gamma_0$, $y_2 = \theta/\theta_0$, and
\begin{equation}
   R_{da} = \left( {E\over \pi A c^2 \theta_0^2 \Gamma_0^2} \right)^{1/(3-s)} \;,
\label{def}
\end{equation}
is the deacceleration radius. $E=M_0\Gamma_0$ is the energy in the
explosion and $\theta_0$ is the initial half-opening angle of the jet. 
The above equations show that for the wind and the uniform ISM models 
$\Gamma\propto t_{obs}^{-1/4}$ \& $t_{obs}^{-3/8}$, respectively, as long as 
$\Gamma\gg\theta_0^{-1}$, where $t_{obs} = \int dt (1-v)$ is the observer time,
$t$ being the lab frame time and $v$ the jet velocity in units of $c$.

Equations (\ref{dynamic1}) and (\ref{dynamic2}) are solved, subject 
to the boundary conditions $y_1=y_2=1$ for $x\ll1$, to determine 
$\Gamma$ and $\theta$ as functions of $r$. For a relativistic jet
with $\Theta=\theta$, i.e. fluid velocity in the radial direction, 
equation (\ref{dynamic2}) reduces to
\begin{equation}
 {d y_1\over dx} = -{x^{2-s} y_1^3 y_2^2\over 2-y_1}, \quad\quad
   {d y_2\over dx} = {1\over f (\theta_0\Gamma_0) x y_1}.
\label{dyna}
\end{equation}
The solution of the equations (\ref{dynamic1}) and (\ref{dynamic2}) is a 
two-parameter family of functions, however in the relativistic case
the solution depends only on the product $\theta_0\Gamma_0$.

One can solve equation (\ref{dyna}) approximately, ignoring the very early time 
behavior, to determine the time when the sideways expansion alters significantly
the jet dynamics. The two relations in equation (\ref{dyna}) can be 
combined to yield a first order differential equation for $y_1 y_2 \equiv y$ 
which is given by
\begin{equation}
{dy\over d\xi} \approx -{y^3\over 2} + {1\over \eta\xi},
\label{dyaaprox}
\end{equation}
with $\eta=f(3-s)(\theta_0\Gamma_0)$, a constant, and $\xi=x^{3-s}/(3-s)$.
An approximate solution to this equation is
\begin{equation}
   y \approx {1\over 2\xi^{1/2}} + \left( {2\over \eta\xi}\right)^{1/3}.
\label{sol}
\end{equation}
Thus, $y\propto\Gamma\theta$ decreases monotonically with radius or time.
The transition to jet sideways expansion starts when the two terms in the
above equation become equal, i.e. $\xi\sim (\eta/16)^2$, and lasts for
an interval in $\xi$ for which $y$ decreases by a factor of $\sim 3$, 
or $x$ increases by a factor of $\sim 3^{3/(3-s)}$. The Lorentz factor 
continues to fall during the transition by a factor of a few. Therefore,
the transition time divided by the time at the start of the transition 
(in observer frame), during which $\alpha_1 \equiv -d\ln(\Gamma-1)/d\ln 
t_{obs}$ increases from $(3-s)/(8-2s)$ to approximately $1/2$, 
is approximately 9x3$^{3/(3-s)}$.
The solution to $y_1$ and $y_2$ can be obtained by inserting the expression 
for $y$ into equation (\ref{dyna}). However, $y_1$ and $y_2$ determined
this way have much larger error than $y$ and should not be used for any
serious calculation.

We solve equations (\ref{dynamic1}) and (\ref{dynamic2}) numerically and 
show the results for $x(t_{obs})$ and $\alpha_1(t_{obs})$ in Figure 1.
Note that the change to $\alpha_1$ from one asymptotic value, corresponding to 
spherical shell expansion, to another, when sideways expansion is well underway, 
takes a long time; the ratio of the final to the initial time for a change in 
$\alpha_1$ of 0.1 for a uniform ISM is $\sim 10^2$ whereas for $s=2$ the ratio 
is $10^3$. For the parameters chosen here $\alpha_1=0.5$ when $\Gamma$ is of 
order a few. In the non-relativistic phase of the jet expansion $\alpha_1=1.2$, 
as for a Sedov-Taylor spherical shock wave.

\section{Synchrotron Emission from Relativistic Jets}
\label{synchrotron}

The synchrotron spectrum in the co-moving frame is taken to be
a sequence of power-laws with breaks at the self-absorption, 
synchrotron peak, and cooling frequencies, as presented in 
Sari, Narayan \& Piran (1998); these frequencies can be found in 
eg. Panaitescu \& Kumar (2000).
All of our numerical results, unless otherwise stated, are obtained
by integrating emission over equal arrival time surface.
Ignoring the radial structure of the jet, the flux received by an
observer located on the jet axis is given by 
\begin{equation}
 f_{\nu} (t_{obs}) = \frac{1}{8\pi d^2} \int_{r_{min}}^{r_{max}}
    \frac{P'_{\nu'}(r)}{\gamma^3 [1 - v\cos\psi(r,t_{obs})]^2} \frac{dr}{r} \;,
\label{fnu}
\end{equation}
where $P'_{\nu'}$ is the co-moving power per frequency at $\nu' = 
\gamma (1 - v\cos\psi) \nu$, $r\cos\psi = ct - ct_{obs}$
and $r_{min}$ and $r_{max}$ are solutions of
\begin{equation}
c t(r_{max}) - r_{max} = c t(r_{min}) - r_{min} \cos\theta(r_{min}) = t_{obs} \; .
\end{equation}

We ignore the angular integration when discussing the analytical calculation
of the observed flux and its power-law decline with time. The observed 
flux at a frequency that is greater than both the cooling frequency, 
$\nu_c$, and the synchrotron peak, $\nu_m$, is proportional to 
\begin{equation}
 f_\nu \propto t_{obs}^{\frac{1}{2}(4-s)-\frac{1}{4}sp} 
    \Gamma^{\frac{1}{2}(p+2)(4-s)} \min\bigl\{(\theta_0 \Gamma_0)^{-2}, y^2\bigr\}.
\label{fnua}
\end{equation} 
At early times when $\Gamma\gg\theta^{-1}$ and $\Gamma\propto
t_{obs}^{-(3-s)/(8-2s)}$, the flux decays as $t_{obs}^{-(3p-2)/4}$.
At late times when $\Gamma\theta\lta 1$ the power-law index for the flux
$\beta \equiv -d\ln f_\nu/d\ln t_{obs} = (4-s)[\alpha_1(p+2)-1]/2 + sp/4 
+ \alpha_2$, where $\alpha_2 \equiv -2d\ln y/d\ln t_{obs}$.

There are two effects that determine the evolution of $\beta$.  One of them, 
the {\it edge effect}, is purely geometrical and results from the angular opening 
$\sim\Gamma^{-1}$ of the relativistic observing cone becoming larger than the jet 
opening angle $\theta$, i.e. the observer ``sees" the edge of the jet. 
The increase to $\beta$ resulting from it is $\alpha_2\lta (3-s)/(4-s)$; 
$\alpha_2$ decreases with time and therefore the jump in $\beta$ is smaller 
for larger $\theta_0$. 
The dimensionless time for $\beta$ to increase by $\alpha_2$ depends 
on the angular position of the observer w.r.t. the jet axis and is 
approximately the ratio of the time when the observer sees the far edge
of the jet to the time when the near side of the jet becomes 
visible. This time is given by
\begin{equation}
R_{t_e} \sim \left[ {\theta_0+\phi_0\over \theta_0-\phi_0}
   \right]^{(8-2s)/(3-s)} = \left[ {1 + P_{\phi_0}^{1/2}\over
    1 - P_{\phi_0}^{1/2} }\right]^{(8-2s)/(3-s)},
\label{rte}
\end{equation}
where $P_{\phi_0}$ is the probability that the observer lies
within an angle $\phi_0$ of the jet axis. For $P_{\phi_0}=0.25$,
$R_{t_e}$ is 18.7 (81) for $s=0$ (2), and during this
time $\beta$ increases by approximately 0.7 (0.4). The dependence of
$R_{t_e}$ on $\phi_0$ becomes much weaker when the emission is
integrated over equal arrival time surface (Figure 2).
This is because the effect of angular integration is to smear the jet-edge
by an angle $1/\Gamma\sim\theta_0/2$, which sets the minimum value of $R_{t_e}$
to be about 10 (10$^2$) for uniform (wind) models.

The other effect which leads to a steepening of the afterglow decay is
dynamical and is caused by the lateral spreading of the jet. During the 
relativistic phase the increase to $\beta$ from the sideways expansion is 
~$\delta\beta = (p+2)(4-s)\delta\alpha_1/2+\delta\alpha_2$; $\delta\alpha_1$ 
and $\delta\alpha_2$ can be read from Figure 1. Since $\alpha_1$ does 
not asymptote to 0.5 so $\beta\not=p$ during the relativistic sideways 
expansion of the jet. \footnote{It should be noted that the 
asymptotic behavior $\beta\rightarrow p$ for $s=0$ (Rhoads 1999) 
is achieved only for extremely narrow jets ($\theta_0 \lta 1^o$), 
so that the jet remains relativistic for a sufficiently long time 
after it starts expanding sideways. It nevertheless serves as a useful, 
quick way of estimating $p$ approximately from the late time light-curve, 
when $\beta$ is no longer increasing.} The value of $\beta$ does, 
however, approach $p$ because $\delta\alpha_1\approx 1/(8-2s)$ 
sometime before the jet becomes non-relativistic and 
$\alpha_2\approx 0$ at this time, thereby giving $\beta\approx p$ 
(see eq. [\ref{fnua}] and Figures 1 \& 2); $\beta$ can exceed $p$, 
as can be seen in Figure 2, however the decrease in $\alpha_2$ 
during the mildly relativistic phase prevents $\beta$ from getting 
much larger than $p$. This result can be extended to any observing 
frequency $\nu > \nu_m$ after an appropriate modification of 
equation (\ref{fnua}). For instance, to consider the case of 
$\nu_c > \nu > \nu_m$ the right side of the equation should be 
multiplied by a factor of $(t_{obs}\Gamma^2)^{(1-3s/4)}$, which 
has little effect on the evolution of $\beta$. The time scale for 
the increase in $\beta$ due to sideways expansion is of order $10^2$ 
($10^3$) for s=0 (2) (see Figure 2). Therefore this effect is smaller 
than that resulting from seeing the jet edge, and it extends over a much 
longer time.

To conclude, we wish to emphasize that for most jets propagating in 
a uniform ISM we are likely to see an increase to $\beta$ of only 
0.6--0.9; the remainder of the increase takes place on a long time 
scale, and thus is hard to detect. For jets in a {\it windy 
medium}, $s=2$, $\beta$ changes by less than about $0.5$ and the 
transition time $R_{t_e}\sim 10^3$. Such a gradual increase to the 
afterglow light-curve power-law index is extremely difficult to 
detect (see Figure 2). For instance, if the edge of the jet becomes 
visible at $t_{obs}\sim 1$ day, the difference in the optical flux at 
the end of 10 days with and without jet is $\sim 0.25$ mag, which can 
be easily missed. Thus, the GRBs studied by Chevalier and Li (1999), 
which show evidence for the {\it wind model}, could in fact have had 
a collimated ejection of material.

\subsection{The Afterglow of GRB 990510}
\label{observations}

The optical emission of the afterglow of GRB 990510 was measured in the V, R and I 
bands between 0.15 and 7 days after the burst and showed the power-law index of the
light-curve, $\beta$, to have increased from $0.82 \pm 0.02$ to $2.18 \pm 0.05$ 
(Harrison et al. 1999) or from $0.76 \pm 0.01$ to $2.40 \pm 0.02$ 
(Stanek et al. 1999) during a dimensionless time $R_{t_e}\approx 30$ which, 
as described previously, 
is not possible to obtain through the effects of the jet sideways expansion alone.
Therefore there must be some contribution to the light-curve steepening due to
the passage of one (or both) of the spectral breaks: the synchrotron peak $\nu_m$
and the cooling frequency $\nu_c$. 

In Figure 3 we show a comparison between the light-curves of GRB 990510 
in the $V$, $R$, $I$ bands and the 8.7 GHz radio data, with a 
model where the cooling frequency $\nu_c$ crosses the 
optical band at $t_{obs} \sim 1$ day. The steepening of the light-curve
has little dependence on the observing band because the ratio of the 
largest to the smallest optical wavelength is 
$\sim 1.5$. Moreover, the integration over angle spreads in time the 
steepening of $\beta$, making it nearly achromatic. An increase of 
$\beta$ by $\sim0.8$ is caused by the jet edge and the sideways expansion, 
and an increase of 0.25 results from the passage of $\nu_c$ through
the observing band. A further increase of $\beta$ of $\sim 0.15$ is 
caused by the passage of $\nu_m$ through the observing band at 
$t_{obs} \sim 0.03$ day (see lower panel of Figure 3); the transition
time for $\beta$ to increase by $\sim(3p-1)/4$ due to the $\nu_m$
crossing is about a decade in the observer frame as a result
of integration over equal arrival time surface, hence one should 
be careful in deducing $p$ from $\beta$ at early times.
All these together give rise to a light-curve that is consistent with 
the data. The model is also consistent with the HST $V$-band observation 
carried out at about a month after the burst (Fruchter et al 1999).
The parameters for the fit are given in the caption for fig. 3 which
yields the energy in the burst to be 2x10$^{49}$ erg. Correcting
for the radiative losses the energy in the burst increases by a
factor of a few to $\lta 10^{50}$ erg. We estimate the uncertainty 
in model parameters
by varying them in such a way that the numerically calculated light-curve 
lies within 3-$\sigma$ of the observed data points. We find the uncertainty 
in the jet opening angle and the burst energy to be a factor of two, 
and $\epsilon_e$, $n$ and $\epsilon_B$ are found to be uncertain by factors
of about 4, 40 and 7 respectively; we note that the radio observations are 
very important in constraining the model parameters.
The electron index $p$ is constrained by the observed $\beta$ before 
and after the $\sim 1$ day break; the error in $p$ is $\sim 5\%$.

The optical emission of the afterglow of GRB 990510 can also be explained by a
model where the synchrotron peak frequency crosses the observed band 
at $\sim 0.1$ day. Its effect on $\beta$ persists for up to $\sim 1$ 
day and yields an increase of $\beta$ of $\sim 0.5$ during the 
early observations. The parameters for the second model differ from 
the one described above (see Figure 3) somewhat. In particular, $\epsilon_e$ 
is larger by a factor of two, the energy per solid angle is smaller by 
a factor of two, and $\theta_0$ is $\sim 20\%$ larger.

\section{Conclusions}

One of the main results of this work is to show that afterglows from
well collimated Gamma-Ray Burst remnants going off in a medium with density 
decreasing as $r^{-2}$ show little evidence for light-curve steepening 
due to jet edge and sideways expansion. This could explain the lack of breaks
in the afterglows of GRB 980326 and GRB 980519, which Chevalier \& Li (1999) 
found to offer support for the {\it wind model}.
Jets can perhaps be detected by the measurement of time dependent polarization.

In a collimated outflow the sharpest break in the light-curve
is produced in a uniform density circum-stellar medium, and is associated 
with the edge of the jet coming within the relativistic beaming cone 
(the {\it edge effect}). The magnitude of this break is $\sim0.7$ 
(0.4) for a uniform ISM (wind model) and occurs over about 1 decade 
(2 decades) in time. Further steepening of the light-curve, associated 
with the sideways expansion of the jet, occurs on a much 
longer time scale of $R_{t_e}\sim$10$^2$ (10$^4$), i.e. weeks to months.

The power-law index for the light-curve of GRB 990510 increased between
days 0.8 and 3 by about 1.35. This is too large and too fast to result 
from jet edge \& sideways expansion effects. However, the observations can be 
explained if either the cooling or the synchrotron peak frequency 
passed through the observing band at about 1 or 0.1 day,
respectively. Models that are consistent both the optical and radio 
data of this afterglow have an opening angle of $\sim 5^o$ and 
energy in the explosion is $\lta 10^{50}$ erg (see Figure 3).

For the afterglow of GRB 990123 the power-law index of the light-curve 
increased by 0.55 between days 1.5 and 3, which can be explained by 
the {\it edge effect} alone (\Meszaros \& Rees 1999). 

\acknowledgments{We thank Peter \Meszaros and Vahe Petrosian for 
useful discussions.}

\begin{figure}
\vspace*{-7cm}
\centerline{\psfig{figure=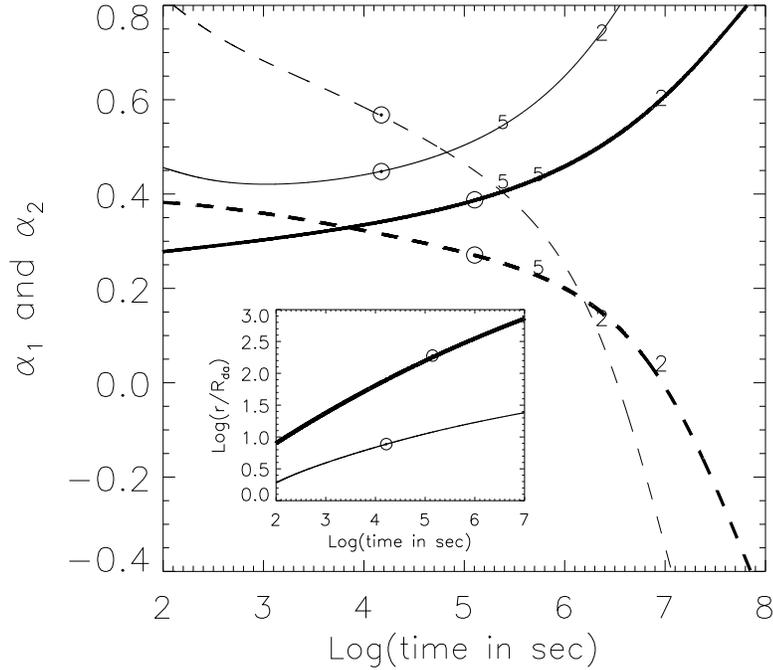,width=10cm}}
\vspace*{1cm}
\figcaption{The main panel shows $\alpha_1=-d\ln(\Gamma-1)/d\ln t_{obs}$
for uniform ISM $s=0$ (thin continuous curve) and the wind model $s=2$
(thick continuous curve), and $\alpha_2=-2d\ln(\theta\Gamma)/d\ln t_{obs}$
for s=0 (thin dashed curve) and s=2 (thick dashed curve). The inset shows 
the evolution of the jet radius. The symbol $\odot$ marks the time when 
$\theta\Gamma=1$, and the symbols 5 (2) denote the time when $\gamma=5$ (2).
For this calculation we took the energy per unit solid angle to be
$3\times10^{53}$ erg sr$^{-1}$ and $\theta_0 = 1/30$ rad. The  density of the
uniform ISM is $1\; {\rm cm}^{-3}$, $A = 5 \times 10^{11}\; {\rm g\, cm^{-1}}$ 
for $s=2$, and $\Gamma_0=300$ (i.e. $\Gamma_0 \theta_0 = 10$).}
\end{figure}
 
\begin{figure}
\vspace*{-4cm}
\centerline{\psfig{figure=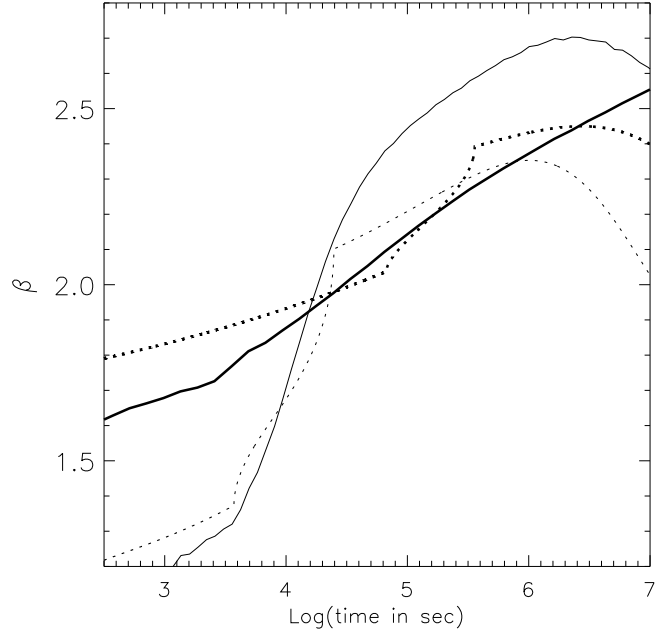,width=10cm}}
\vspace*{1cm}
\figcaption{The panel shows $\beta\equiv -d\ln f_\nu/\ln t_{obs}$ 
for $\nu>\max(\nu_m,\nu_c)$ and $\phi_0=\theta_0/\sqrt{5}$.
The thin \& thick dashed lines are for $s=0$ \& 2 respectively;
the observed flux in these cases was calculated without proper
angular integration over the jet surface. The sharp increase to 
the value of $\beta$ seen in the dashed curves arises from 
the {\it edge effect} described in the text.
The thin and the thick continuous curves are for $s=0$ \& 2 respectively
and these calculations included integration over equal arrival time
surface; note that this smoothes out sharp changes in $\beta$.
For all of these calculations we took the energy per unit solid angle 
to be $3\times10^{53}$ erg sr$^{-1}$, $\theta_0=1/30$ rad 
($\theta_0\Gamma_0=10$), $p=2.5$, the density of the uniform ISM 
is 1.0 cm$^{-3}$ and $A = 5 \times 10^{11}\; {\rm g\, cm^{-1}}$ for $s=2$.
The results shown here are independent on the sideways expansion speed (i.e.
the parameter $f$ in eq. [\ref{theta}]), as long as this speed is relativistic.}
\end{figure}

\begin{figure}
\vspace*{-1cm}
\plotone{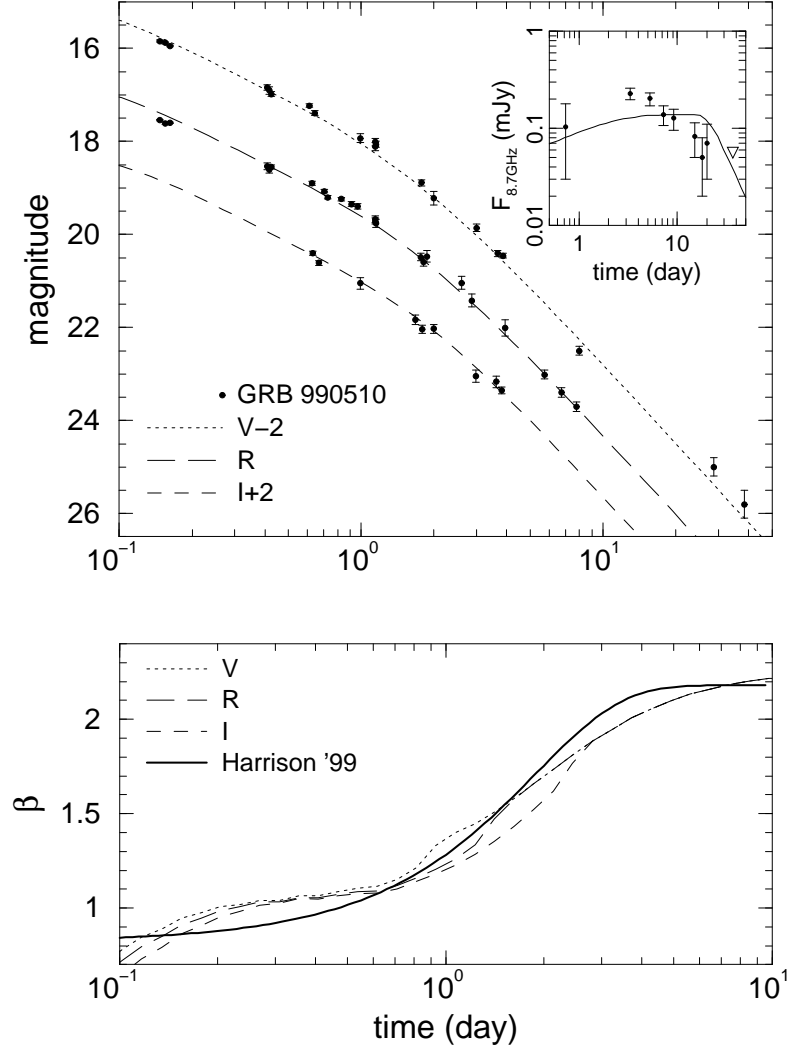}
\vspace*{-5cm}
\figcaption{
Comparison between the observed and the theoretically calculated
light-curves for the afterglow of GRB 990510 in the radio and the
optical bands. Most of the data is taken from Harrison et al. (1999); 
the optical data was supplemented with observations reported in GCN 
 Circulars (http://gcn.gsfc.nasa.gov/gcn/gcn3) by Beuermann et al. 1999, 
Marconi et al. 1999, Pietrzynski \& Udalski 1999; the latest measurements 
in the $V$-band were taken with the HST (Fruchter et al. 1999). 
 The model light-curves are calculated for an observer located on the jet 
axis, which gives the fastest decline of the light-curve. The jet has 
energy per solid angle $E/(\pi\theta_0^2) = 1.2\times 10^{52}/4\pi$ erg 
sr$^{-1}$ and a half-angle $\theta_0=0.085$ rad. The electrons acquire 
$\epsilon_e=0.3$ of the internal energy after shock acceleration, 
 the magnetic field energy is $\epsilon_B=0.04$ of that of the shocked 
gas, and the electron index is $p=2.2$ . The external medium is 
homogeneous with $n=0.23\;{\rm cm^{-3}}$. The redshift of the source 
is $z=1.62$. The cooling frequency passes through the observing 
 window at $t_{obs}=1.2$ day steepening the afterglow light-curve 
while the sideways expansion is effective. The inset of the upper 
panel shows the 8.7 GHz emission. The lower panel shows a comparison of 
the numerically computed power-law index ($\beta$) for the decline of the 
afterglow of GRB 990510 and the observed one, as obtained by the fitting 
formula given in Harrison et al. (1999).
}
\end{figure}

\end{document}